\documentclass[sigconf]{acmart}

\AtBeginDocument{%
  \providecommand\BibTeX{{%
    \normalfont B\kern-0.5em{\scshape i\kern-0.25em b}\kern-0.8em\TeX}}}

\setcopyright{rightsretained}
\copyrightyear{2020}
\acmYear{2020}
\setcopyright{rightsretained}
\acmConference[FAT* '20]{Conference on Fairness, Accountability, and Transparency}{January 27--30, 2020}{Barcelona, Spain}
\acmBooktitle{Conference on Fairness, Accountability, and Transparency (FAT* '20), January 27--30, 2020, Barcelona, Spain}
\acmDOI{10.1145/3351095.3372826}
\acmISBN{978-1-4503-6936-7/20/02}

\usepackage{tabularx}
\usepackage{xcolor}
\usepackage{ulem}

\newcolumntype{L}[1]{>{\raggedleft\arraybackslash}p{#1}}

\begin{document}

\title{Towards a Critical Race Methodology in Algorithmic Fairness}

\author{Alex Hanna}\authornote{Both authors contributed equally to this research.}
\author{Emily Denton}\authornotemark[1]
\author{Andrew Smart}
\author{Jamila Smith-Loud}
\affiliation{\{alexhanna,dentone,andrewsmart,jsmithloud\}@google.com}

\renewcommand{\shortauthors}{Hanna et al.}

\begin{abstract}
We examine the way race and racial categories are adopted in algorithmic fairness frameworks. Current methodologies fail to adequately account for the socially constructed nature of race, instead adopting a conceptualization of race as a fixed attribute. Treating race as an attribute, rather than a structural, institutional, and relational phenomenon, can serve to minimize the structural aspects of algorithmic unfairness. In this work, we focus on the history of racial categories and turn to critical race theory and sociological work on race and ethnicity to ground conceptualizations of race for fairness research, drawing on lessons from public health, biomedical research, and social survey research. We argue that algorithmic fairness researchers need to take into account the multidimensionality of race, take seriously the processes of conceptualizing and operationalizing race, focus on social processes which produce racial inequality, and consider perspectives of those most affected by sociotechnical systems.

\end{abstract}

\ccsdesc[500]{Applied computing~Sociology}
\ccsdesc[300]{Social and professional topics~Race and ethnicity}

\keywords{algorithmic fairness, critical race theory, race and ethnicity}


\maketitle

\vspace{75pt}

\begin{quote}
    The problem does not end with the collection of racial data; it only begins. The problem accelerates when we attempt to analyze these data statistically... The racialization of data is an artifact of both the struggles to preserve and to destroy racial stratification. Before the data can be deracialized, we must deracialize the social circumstances that have created racial stratification.

    -- Tufuku Zuberi \cite[pp. 102]{zuberi2001thicker}
\end{quote}

\section{Introduction}
In recent years, there has  been increasing recognition of the potential for algorithmic systems to reproduce or amplify existing social inequities. In response, the research field of algorithmic fairness has emerged. This rapidly evolving research area is focused on developing tools and techniques with aspirations to make development, use, and resulting impact of algorithmic systems conform to various social and legal notions of fairness. The concept of fairness, in addition to being situational, evolving, and contested from a number of philosophical and legal traditions, can only be understood in reference to the different social groups that constitute the organization of society. Consequently, the vast majority of algorithmic fairness frameworks are specified with reference to these social groups, often requiring a formal encoding of the groups into the dataset and/or algorithm. 

However, most social groups relevant to fairness analysis reflect highly contextual and unstable social constructs.  These social groups are often defined with recourse to legal anti-discrimination concepts such as "protected classes," which, in the US, refers to race, color, national origin, religion, sex, age, or disability.  However, the process of drawing boundaries around distinct social groups for fairness research is fraught; the construction of categories has a long history of political struggle and legal argumentation.

Numerous recent works have highlighted the limitations of current algorithmic fairness frameworks \cite{hoffmann2019fairness, Selbst2019}. Several of these critiques point to the tendency to abstract away critical social and historical contexts and minimize the structural conditions that underpin problems of algorithmic unfairness. We build on this critical research, focusing specifically on the use of race and racial categories within this field. IN this literature, the topic of the instability of racial categories has gone relative unexplored, with the notable exception of Benthall and Haynes \cite{Benthall2019}, which we discuss in detail below.

Race is a major axis around which algorithmic allocation of resources and representation is bound. It may indeed be the most significant axis, given attention by investigative journalists (e.g. \cite{propublica}) and critical race and technology scholars (e.g. \cite{noble2018algorithms, browne2015dark, benjamin2019race, chun2009, brock2009life}). Because of this, it is imperative that the race-based methodologies and racial categories themselves are interrogated and critically evaluated. 

In this paper we develop several lines of critique directed at the treatment of race within algorithmic fairness methodologies. In short, we observe that current methodologies fail to adequately account for the socially constructed nature of race, instead adopting a conceptualization of race as a fixed attribute. This manifests in the widespread use of racial categories as if they represent natural and objective differences between groups. Treating race as an attribute, rather than a structural, institutional, and relational phenomenon, in turn serves to minimize the structural aspects of algorithmic unfairness. 

The process of operationalizing race is fundamentally a project of racial classification and thus must be understood as a political project, or, more specifically, what Omi and Winant \cite{omi2014racial} refer to as a racial project. The tools for measuring individual characteristics, such as national censuses and other population-level survey instruments, were and are still often based in the politics of racial oppression and domination. While we acknowledge the importance of measuring race for the purposes of understanding patterns of differential performance or differentially adverse impact of algorithmic systems, in this work, we emphasize that data collection and annotation efforts must be grounded in the social and historical contexts of racial classification and racial category formation. To oversimplify is to do violence, or even more, to re-inscribe violence on communities that already experience structural violence.

It is useful to distinguish between two ways in which race comes into play in algorithmic fairness research: (i) the operationalization of race, i.e. the process of converting the abstract concept of race into something that is concrete and measurable and (ii) the use of racial variables within algorithmic frameworks. These aspects are tightly interconnected since race must be operationalized before racial variables can be utilized. 

Our contributions are as follows: we review the use of race in prior fairness work and discuss an intervention in the debate on racial categories. We then survey the history of racial classification, scientific racism, and racial classification in national censuses. We introduce the notion of multidimensionality of race, and subsequently review how different disciplines have dealt with the tricky problem of operationalizing race, most notably in biomedical, survey, and public health research. In light of these discussions, we address how race has been treated in the group fairness and disaggregated analysis paradigms. Lastly, we argue that fairness researchers need to take into account the multidimensionality of race, take seriously the processes of conceptualizing and operationalizing race, focus on processes of racism, and consider perspectives of those most affected by sociotechnical systems.
\section{The problem with racial categories}

In perhaps the best-known debate surrounding algorithmic fairness, Julia Angwin and her colleagues illustrated the differing rates at which Black defendants had been issued high pre-trial risk assessment scores by the COMPAS recidivism prediction algorithm compared to white defendants \cite{propublica}. Equivant -- then, Northpointe, the company which developed COMPAS -- defended the system by reanalyzing the ProPublica data \cite{dieterich2016compas} and ProPublica later responded \cite{angwin2016j}. The debate has become a cornerstone of algorithmic fairness research; to date, the original story has some 700 citations on Google Scholar.

The Propublica analysis of COMPAS, as well as the responses from Northpointe and other secondary analyses performed by third-party researchers, relied on data from the Broward County Sheriff's Office which was obtained through a public records request. Race classifications in this data identified defendants as Black, White, Hispanic, Asian, Native American or Other. These categories look familiar -- they are nearly identical to the way that the US Census defines race. However, there are the notable absences of the categories "Native Hawaiian or Other Pacific Islander", and there is a redefinition of "Hispanic" as a race rather than an ethnicity.

In the methodological appendix to the original article \cite{larson2016we}, Jeff Larson and other ProPublica authors do not delve into how the race of the defendant is measured. Larson admits that he did not know how Broward County classified individuals \footnote{Personal correspondence, Jeff Larson, August 19, 2019.}. It's not clear why Broward County used the modified Census racial schema, but, as detailed below, it may have to do with a 1977 directive from the federal US Office of Management and Budget. In general, approaches to measuring race for police records are inconsistent across municipalities. They can rely on self-identification, state records, or observation by criminal justice workers. In her work on racial disparities in incarceration in Wisconsin, sociologist Pamela Oliver notes that the race of a single individual can change across records even within a single jurisdiction \footnote{Personal correspondence, Pamela Oliver, August 19, 2019.}. Therefore, even in the most famous debate of the field, we don't know why the data take on a particular racial schema, nor do we have information about how defendants are racially categorized.

\subsection{Using race-like categories?}

While an area of major concern for critical race and technology scholarship, the use of racial categories in algorithmic fairness research (i.e. the research community which has emerged around venues like FAT* and AIES) has largely gone unquestioned. There are two important exceptions. First, in the original Gender Shades paper which helped establish intersectional testing, Buolamwini and Gebru note the instability of race and ethnicity labels, which justifies their use of skin tone \cite{buolamwini2018gender}. Second, the critique levied by computational social scientist Sebastian Benthall and critical race scholar Bruce D. Haynes \cite{Benthall2019} describes how little attention has been given to the meaning of protected class labels, the political orientation of class labels, and the normative assumptions of machine learning design. They focus on the typical referent group -- African-Americans in the US -- and discuss how "racial classification is embedded in state institutions, and reinforced in civil society in ways that are relevant to the design of machine learning systems." Race is invoked because "racial classification signifies social, economic, and political inequities anchored in state and civic institutional practices." 

We wholly agree with their overall argument, which is that algorithmic fairness research has been largely silent on the issues of racial statistics and categories, and that in that silence, has thus reified them. In their criticism of the COMPAS debate, we echo their argument:

\begin{quote}
    rather than taking racial statistics... at face value, the process that generates them and the process through which they are interpreted should be analyzed with the same rigor and skepticism as the recidivism prediction algorithm. Thematically, we argue that racial bias is far more likely to come from human judgments in data generation and interpretation than from an algorithmic model, and that this has broad implications for fairness in machine learning \cite[pp. 291]{Benthall2019}.
\end{quote}

We diverge from their critique, not necessarily in their problem formation, but in their solution. They offer up a narrow path to follow which plots a course between the Scylla of "[m]achines that ... [reify] racial categories that are inherent unfair" and the Charybdis of "systems that allocate resources in ways that are blind to race [which] will reproduce racial inequality in society" \cite[pp. 294]{Benthall2019}. In their solution, they propose using an unsupervised machine learning method to create race-\textit{like} categories which aim to address "historical racial segregation with reproducing the political construction of racial categories." One virtue of this method is that it is attentive to creating a metric which makes sense with respect to a particular social inequality, that is, spatial segregation. This is something we agree with in our adoption of a multidimensional perspective below.

We'd like to raise a few complexities in their formulation which make us hesitant to adopt their solution. First, it would be a grave error to supplant the existing categories of race with race-like categories inferred by unsupervised learning methods. Despite the risk of reifying the socially constructed idea called race, race does exist in the world, as a way of mental sorting, as a discourse which is adopted, as a social thing which has both structural and ideological components. In other words, although race is social constructed, race still has power. To supplant race with race-like categories for the purposes of measurement sidesteps the problem. 

Second, supplanting race with race-like categories depends highly on context, namely how race operates within particular systems of inequality and domination. Benthall and Haynes restrict their analysis to that of spatial segregation, which is to be sure, an important and active research area and subject of significant policy discussion (e.g. \cite{massey1993american, rugh2010racial}). However, that metric may appear illegible to analyses pertaining to other racialized institutions, such as the criminal justice system, education, or employment (although one can readily see their connections and interdependencies). The way that race matters or pertains to particular types of structural inequality depends on that context and requires its own modes of operationalization.

Third, and relatedly, as Wendy Chun and Ruha Benjamin have discussed \cite{chun2009, benjamin2019race}, race operates both \textit{with} and \textit{as} technology. At the same time we focus on the ontological aspects of race (what is race, how is it constituted and imagined in the world), it is necessary to pay attention to what we do with race and measures which may be interpreted as race. The creation of metrics and indicators which are race-like will still be interpreted as race. As we discuss more below, the example of genomics research is indicative. Even as genomics researchers warn not to interpret genetic ancestry as race or ethnicity \cite{Yudell564, bolnick2007science}, this has not prevented people (e.g. customers of direct-to-consumer genetic testing companies) from interpreting them as such \cite{Roth2018}.

Finally, we'd like to underline the {\it infrastructural} criticism embedded in the act of categorization itself. By infrastructure, we refer to the way that classifications and standards form the infrastructure of our existing information society \cite{sorting}. Inverting infrastructures and seeking their fissures allows us to understand how they are used in practice and how politics are embedded in their creation. This point dovetails with the critiques levied by abolitionist activists, scholars, and technologists \cite{barabas2017interventions, Abdurahman2019, benjamin2019race}. J. Khadijah Abdurahman \cite{Abdurahman2019} argues that to study algorithmic fairness is to sidestep the problem beyond the algorithmic frame, that we, with communities who are disproportionally affected by redlining, predictive policing, and surveillance, should not just contest how "protected classes within algorithms are generated-but [should] viscerally reject the notion human beings should be placed in cages in the first place." The task of using racial classifications at all requires us to expand the frame to consider what the larger implications of classifying are. Who is doing the classifying? For what purpose are they classifying and to what end?
\section{Histories of racial categorization}

We recount histories of racial category construction by administrative and scientific bodies as a means of highlighting the decidedly social constructivist notion of race. As Hacking \cite{hacking1999social} has said, social constructivism is the act of "making up people." Social construction does not mean that things in the world are not "real" \cite{elder2012towards}, but that that thing -- be it a racial, gender, or class category -- was brought into existence or shaped by historical events, social forces, political power, and/or colonial conquest, all of which could have been very different.  When we demonstrate that something is socially constructed, it becomes clear that it could be constructed differently, and then we can start to demand changes in it \cite[pp. 6-7]{hacking1999social}.

Race as a concept is widely acknowledged by the social sciences to be a social construction. In the racial formation framework, Omi and Winant \cite{omi2014racial} discuss how race is both real and socially constructed. The social constructedness of race decenters the concept as a property of individuals determined by phenotypical properties. Instead, social constructivism places race within specific history and context. The constructedness of race in any given point in time is tied to the specific \textit{racial project}. A racial project is "simultaneously an interpretation, representation, or explanation of racial identities and meanings, and an effort to organize and distribute resources (economic, political, cultural) along particular racial lines" \cite[pp. 56]{omi2014racial}. Those racial projects can range in scale and kind, in the microinteractional to the structural. 

Race is not something that arrived whole cloth, but needed to be made up. Race, although socially constructed, of course has very salient material effects. Accordingly, the most modern understanding of race argues that it is inauthentic to conceptualize race as a natural property of individuals. The "naturalization" of race and racial categories do not come from nowhere. It has been constructed and been continually reproduced as part of a project of upholding a particular type of racial project -- one that is historically and culturally bound. Historian Ibram X. Kendi, for instance, thoroughly outlines the history of race and racial projects from 15-century Europe to the present \cite{kendi2017stamped}. Projects that "misrecognize" race as natural \cite{desmond2009racial} are driven by the hegemonic nature of racial projects \cite{omi2014racial}. It would be more accurate to describe race as relational and as a property of institutions, organizations, and larger structures \cite{ray2019theory}. Race may be more accurately discussed as having relational qualities, with dimensions that are symbolic or based on phenotype, but that are also contingent on specific social and historical contexts. 

\subsection{Classification and the Racial History of Social Statistics}

Classification is a process imbued with social, economic, and organizational imperatives \cite{foucault2005order, sorting}. Those who do the categorizing have their own institutional and occupational priorities. Categories themselves become a type of infrastructure: they are the ground upon which other structural and ideational elements are built. When they work, they are invisible, but when they break down, the boundaries and assumptions begin to show.  With the development of markets and the rise of technological innovation and actuarial science, classification of individuals has moved from the more general assessment based on subgroup distinction to individuation based on market imperatives.

Although in the modern neoliberal era, markets have been an emerging force in categorization and segmentation \cite{fourcade2016seeing}, the entity tasked with categorization above all others has been the state. Almost as soon as humans began living in agriculturally-based communities, it became necessary to count the number of people living under the control of a certain party \cite{scott2017against}. Categorization was also used to streamline counting items of like type -- such as trees in a forest, grain, or human beings. Further, since early grain-based agriculture necessitated vast numbers of essentially slave-laborers, rulers needed to know how many slaves there were and moreover, who counted as a slave. As Scott argues, during the agricultural revolution there was initially strong resistance to transitioning from life as a hunter-gatherer to a more hierarchical, regimented, and predictable existence as a slave harvesting grain and to "being counted" by newly-created states \cite{scott2017against}. 

The modern discipline of statistics grew out of a bureaucratic need to manage large populations and natural resources \cite{herzfeld1993social}. More specifically and perniciously, the field of social statistics emerged from the need to make differentiations between white Europeans and their descendants, and other peoples. Francis Galton was a major figure in the development of social statistics; he was also a eugenicist and a major proponent of using statistics as a means to justify racial superiority of European-origin peoples \cite{Zuberi2000, zuberi2001thicker, bonilla2008toward}. As Galton writes in his \textit{Hereditary Genius}:

\begin{quote}
    The natural ability of which this book mainly treats, is such as a modern European possesses in a much greater average share than men of the lower races. There is nothing either in the history of domestic animals or in that evolution to make us doubt that a race of sane men may be formed, who shall be as much superior mentally and morally to the modern European as the modern European is to the lowest of the Negro races (quoted in \cite{Zuberi2000}).
\end{quote}

The practice of racial classification that emerged in social statistics is closely tied to the project of nation-building, in multiple senses. In the first case, racial classification was a prerequisite to the project of European colonization of the Americas. The racialization of chattel slavery is a modern phenomenon. European colonial projects developed alongside Enlightenment ideas of liberal democracy. Creating structures of racial stratification necessitated ideational components that would legitimate the enslavement of Africans \cite{zuberi2001thicker}. Until the early 20th century  in the West and the rise of cultural anthropologists such as Franz Boas, there was widespread support for eugenics and scientific racism, and the belief that the state must intervene to protect the physical and mental characteristics of the white race. In 1926, 23 of 48 states had laws permitting sterilization \cite{scott1998seeing}. At the height of miscegenation legislation (that is, laws against marriages which were "interracial" or "cross-cultural" in nature), 41 American colonies and states had laws against the practice \cite{pascoe1996miscegenation}. As the project of nation-building (more specifically, colonization of non-European nations) waned in the 1960s, so did the popularity of eugenics as a scientific practice. However, as Syed Mustafa Ali notes, the scientific racism undergirding colonialism has persisted as "`sedimented' ways of knowing and being -- based on systems of categorisation, classification, and taxonomisation and the ways that these are manifested in practices, artefacts and technologies" \cite{ali2016brief}. Technologies of racial classification, such as blood quantum for Native Americans, persist and upon legacies of settler colonialism. These technologies become enshrined in folk understandings of racial percentages, such as genetic ancestry \cite{tallbear2013native}, or in law, as blood quantum remains a legal requisite to tribal membership, tribal constitution, and land claims.

Second, since the founding of the Americas, delineating racial boundaries has been part and parcel of the state-building project. Assessing and evaluating racial and ethnic boundaries is a state practice that coincides not only with a cultural understanding of who belongs to the polity, but has and continually is a prerequisite for formal citizenship in many countries. In the pre-war United States, appeals to citizenship for non-native born residents were fundamentally appeals to whiteness. Lopez documents how appeals to whiteness and the boundaries of the category of whiteness flexed and contracted in legal argumentation \cite{lopez2006white}. The 1790 Naturalization Act restricted citizenship to "any alien, being a free white person" who had been in the United States for at least two years. It wasn't until the passage of the Immigration and Nationality Act of 1952 that the explicit racialized nature of naturalization was restricted, although the Act laid the groundwork for the current restrictive race-based ratio system which exists today. Maghbouleh demonstrates how, prior to 1952, Iranians could be used as a "racial hinge" to argue both for and against claims to whiteness, and how citizenship claims critically depended on who could make this claim \cite{maghbouleh2017limits}. 

Third, national censuses are a critical component of state machinery. State enumeration of populations works to facilitate the administrative and bureaucratic functions of the modern nation-state. As Scott points out:

\begin{quote}
    State simplifications such as maps, censuses, cadastral lists, and standard units of measurement represent techniques for grasping a large and complex reality; in order for officials to be able to comprehend aspects of the ensemble, that complex reality must be reduced to schematic categories. The only way to accomplish this is to reduce an infinite array of detail to a set of categories that will facilitate summary descriptions, comparisons, and aggregation. The invention, elaboration, and deployment of these abstractions represent, as Charles Tilly \cite{tilly1990coercion} has shown, an enormous leap in state capacity -- a move from tribute and indirect rule to taxation and direct rule \cite[pp. 77]{scott1998seeing}.
\end{quote}

Furthermore, censuses not only quantify members of racial social groups, but also \textit{create and constitute the boundaries of these social groups themselves}. Melissa Nobles outlines how racial classifications are taken for granted by policymakers, but how national censuses not only enumerate polity members into racial categories, but also help to reinscribe racial categories themselves, that is, "[c]ensus-taking is one of the institutional mechanisms by which racial boundaries are set" \cite[pp. xi]{snipp2003racial}.  In the US, the national census is the sole basis for which representation is guaranteed in the House of Representatives and the Electoral College. In the antebellum US, the Three-Fifths Compromise, in which Black slaves were counted as three-fifths of a human for the purposes of appropriation and taxation, is the most dramatic example of racial classification in counting populations. Whereas the passage of the 14th Amendment guaranteed all Americans being counted as a whole person regardless of race, racial categories defined in the Census in the latter part of the 19th Century reflected fractional referents to Black racial lineage based in the ascendant movement of eugenics and race science mentioned above. Categories of "octoroon" and "quadroon", along with "mulatto" denoted fractional heredities of the Black population \cite{snipp2003racial}.

Because of the allocative nature of census-taking and the bureaucracies associated with the practice, what goes into the census is politically contested by mobilizing racial and ethnic groups. The political contestation of the census has been laid bare recently with the attempted introduction of a citizenship question into the US Census and its potential for its dampening effects on participation by Latinx groups \cite{liptak2019}. For instance, after the 1950 Census, in light of the discoveries of dramatic undercounts of non-white populations and with rising pressures from the Civil Rights Movement, the 1960 Census shifting from interview identification to asking individuals to self-identify their race. As Snipp rightly points out, this procedural change was a "fundamental redefinition of race." Population statistics formerly based on phenotypical appearances were now based on "cultural affiliation and other deeply held personal considerations beyond the pale of conventional demographic inquiry" \cite[pp. 570]{snipp2003racial}. This shift was not only bureaucratic but also (perhaps unknowingly by the Census Bureau) ontological: race as measured by the Census aligned more with the social constructionism and less with essentialism. The Office of Management and Budget (OMB) Directive No. 15 of 1977 instituted a standard in racial classifications used by government agencies and developed a standard for five distinct groups: (a) American Indians and Alaska Natives, (b) Asian and Pacific Islanders, (c) Non-Hispanic Blacks, (d) Non-Hispanic Whites, and (e) Hispanics. The impact of this directive had far-reaching implications, as this categorization "permeated every level of government, many if not most large corporations, and many other institutions such as schools and nonprofit organizations" (\cite{omi2001changing}, cited in \cite{snipp2003racial}). The 1997 revision of Directive No. 15 separated out Native Hawaiians and other Pacific Islanders, set Hispanics and Latinos apart as an ethnic group (rather than a racial group), and allowed respondents to provide multiple responses to racial heritage \cite{snipp2003racial}. 

Dramatic changes in governance have large downstream effects for the accuracy of census counts and their categories. Khalfani et al. document the differences between the South African numbers before and after the reign of Mandela's African National Congress, and find that the raw counts and the estimations from aerial photography both dramatically undercount the majority African populations, as well as women \cite{khalfani2008race}. The US Census Bureau spends a non-trivial amount of effort getting enumeration correct for marginalized groups; however, those efforts can be stymied with the winds of partisanship and political administration. Even without the issues raised by a potential citizenship question, decreased or flat-lined funding has left the Census Bureau  understaffed and unable to thoroughly test new methodologies \cite{elliott2019assessing}.

In sum, the act of classification cannot be divorced from the group who is doing the classification and the organizational and institutional pressures of performing classification. Racial classification has a long history within quantitative social science, and, in some ways, is the reason that social statistics have developed the way they did. That history is grounded in eugenicist thought and practice, and much of that work provided the infrastructure for census-taking in the US, Canada, Latin America, and South Africa. More expansive categorization resulted from social and political movements in those countries. In short, categorization itself is a technological infrastructure within which institutional racism continues to reside. In this way, we can realize the ways in which race intersects with technology but also as a technology in and of itself. The framing of race as technology shifts the terrain of the discussion -- rather, from what race is and how we can classify individuals, to what we can do with race and how to make it do different (and explicitly anti-racist) things.

\subsection{The multidimensionality of race}

As highlighted by the 1960 Census shift from ascribed to self-identification, classification is largely contingent on the particular appraisal of race used. When we say "race", we may be discussing self-identification, but we also may be referring to phenotypical features or observed assessments from third parties. Racial classifications are uneasily balanced not only on the particular unstable equilibrium of racial projects, but also on the micro-level processes of race appraisals themselves. When it comes to measurement and operationalization, "race" is not a single variable, but many differing and sometimes competing variables. Roth \cite{roth2016multiple} refers to this constellation of variables as the "multiple dimensions" of race. Race manifests in many ways in the real world. Each of these different dimensions can be measured in a different manner for empirical research, and each of these different dimensions have distinct downstream outcomes. It may therefore be more appropriate to study a particular outcome using measurement which reflects a respective dimension of race. Table \ref{tab:table1} reproduces, with minor amendments, the table from \cite{roth2016multiple} which outlines each of these dimensions.

\begin{table*}[]
\centering
\renewcommand{\arraystretch}{1.2}
\begin{tabularx}{\textwidth}{lp{.24\textwidth}p{.25\textwidth}p{.25\textwidth}}

\toprule
Dimension & Description & Typical Measurement & Outcomes it may be appropriate for studying \\ \midrule
Racial identity & Subjective self-identification, not limited by pre-set options. & Open-ended self-identification question & Political mobilization; assimilation; social networks; voting; residential decision-making; attitudes \\
Racial Self-Classification & The race you check on an official form or survey with constrained options (e.g. the Census) & Closed-ended survey question & Demographic change; vital statistics; disease and illness rates \\
Observed Race & The race others believe you to be & Interviewer classification & Discrimination; socioeconomic dispartiies; residential segregation; criminal justice indicators; health care/service provision \\
- Appearance-Based & Observed race based on readily observable characteristics & Interviewer classification with instructions to classify on first observation & Racial profiling; discrimination in public settings \\
- Interaction-Based & Observed race based on characteristics revealed through interaction (e.g. language, accent, surname) & Interview classification with instructions to classify after interaction or survey & Workplace discrimination; housing discrimination; language/accent-based discrimination \\
Reflected Race & The race you believe others assume you to be & A question such as "What race do most people think you are?" & Self-identification processes; perceived discrimination \\
Phenotype & Racial appearance & Usually interviewer classification, but also "objective" characteristics such as: skin color; hair texture and color; nose shape; lip shape; eye color & Discrimination; socioeconomic dispartiies; residential segregation; criminal justice indicators; health care/service provision \\
\bottomrule
\end{tabularx}
\caption{Multiple dimensions of race. Reproduced and amended from \cite{roth2016multiple}. Note that we have excluded "racial ancestry" from this table. Genetics, biomedical researchers, and sociologists of science have criticized the use of "race" to describe genetic ancestry within biomedical research \cite{Harawa2009, neal2008, Yudell564, fujimura2014clines}, while others have criticized the use of direct-to-consumer genetic testing and its implications for racial and ethnic identification \cite{bolnick2007science, tallbear2013native, rajagopalan2012making}.}

\label{tab:table1}
\end{table*}

\textit{Racial identity} and \textit{racial self-classification} refer to subjective self-identification of race. In theory, self-identification is one dimension but in practice these two are measured in different ways and have differing downstream outcomes. Racial identity can be thought of as the sense of self, the identity and broad group with which someone belongs. Accordingly, its measurement instrument is that of an open-ended interviewer or survey question. Racial self-classification refers to the discrete categories that one marks on intake forms, census surveys, and self-identification documents for employment. This dimension, as noted earlier, is highly constrained by the categories determined {\textit a priori} by institutional and organizational directives, such as OMB Directive No. 15.

\textit{Observed race} refers to the race which others believe ascribed to an individual. This may be based upon first impression by an interviewer or annotator. The third-party can make an assessment based solely on \textit{appearance} or \textit{interaction}. Appearance-based observed race depends on readily observable characteristics, whereas interaction-based observed race depends on language, accent, and other physical, aural, or social cues. Reflected race is the race which an individual believes that others ascribe to them. Although this is a first-person evaluation, it is the first-person evaluation of a third-party appraisal. 

\textit{Phenotype} refers to the set of objective characteristics which characterize appearance. A significant literature revolves around skin color appraisals as a means for determining particular social outcomes (e.g. \cite{Telles2015, monk2015cost}) and the use of skin color as a means of evaluating facial recognition systems (e.g. \cite{buolamwini2018gender, Raji2019}). Other race-related phenotypical features include hair texture and color, eye shape and color, and lip shape. The related outcomes for phenotypical features are the same as those for observed race, but are not constrained in measurement by discrete racial categorization\footnote{In a sense, the "observed race/appearance-based" dimension and the phenotype dimension are collapsed for computer vision systems, because a computer vision system can only "see" the objective and phenotypical.}.


The dimensions of race noted here are, of course, unstable across time, place, and context. Just as racial classifications themselves are tied to particular racial projects, modes of self-identification, observed race, and reflected race will be tied to the dynamics of that project. Self-identification has been shown to be responsive to one's arrest records \cite{penner2015disentangling} and to one's class position \cite{Telles1998}. Observed race depends on the interviewer or annotator who is doing the observing. Reflected race depends not only on phenotype but also on the social status of the group in which the individual believes they are being recognized as. These positions shift across time, not only across the life course but also across the \textit{longue dur\'{e}e} of different racial projects. 

Three major implications follow from the of multiple dimensions paradigm. The first is that there may be inconsistencies across different dimensions of race. A single individual can have differing racial appraisals. In the health field, for instance, Roth cites several studies in which individual self-identification differs from medical records, interviewer classifications, or death certificates. Second, the dimensions can cross the boundaries between each other. Phenotype and reflected race can impact self-identification but do not fully determine it.  However, the third and most important implication, also highlighted in Table \ref{tab:table1}, is that different dimensions will be associated with differing outcomes. Therefore, using a measurement of race which does not reflect social processes associated with it may misrepresent the scope of racial disparities at best and underreport them at worst.

\section{Lessons from other disciplines}

The variance in what the concept of race can mean has been noted as "antithetical to the tenets of scientific research, which, in its ideal form, demands that analytical variables be consistent and their categories mutually exclusive" \cite{Lee2009}. Yet, the social centrality of race positions it as a critical concept in many fields of study. In this section, we review how other disciplines, such as public health research, biomedical research, and studies of social inequality, have grappled with the use of racial categories within their disciplines. We surface these discussions to understand how algorithmic fairness research can learn from these methodological approaches.


\subsection{Limitations in operationalizing race}

Scholars from a range of disciplines have observed a widespread lack of clarity around the use of racial variables within their respective fields. Race is often inconsistently conceptualized and measured across studies; definitions and processes of operationalization tend to be insufficiently documented. These inconsistencies pose serious challenges for the validity and utility of research results. For example, inconsistent categories and classification schemes can result in mismatches between different measures of race for the same individual across different time points or across studies. This in turn can significantly impact health statistics \cite{Kaufman1999, Sandefur2004}, population statistics \cite{Bailey2013, Loveman2012}, and measures of social inequality \cite{Saperstein2006}. Country-specific understandings of what "race" references and local racial categorization schemes pose significant challenges for international comparisons \cite{Roth2017}. 

Standardizing racial taxonomies and measurement practices does not sufficiently mitigate the many concerns scholars have raised regarding the use of racial categories in scientific studies. In fact, the widespread uncritical adoption of racial categories, standardized or not, can erode awareness of the social and political histories of racial taxonomies and reify racial categories as natural kinds \cite{Salter2013, Duster1050, smart2008, Fullwiley2007, Fullwiley2008, Mcharek2013, Braun2007}. The reification of race as a natural category can in turn re-entrench systems of racial stratification which give rise to real health and social inequalities between different groups \cite{Kaufman1999, Sewell2016}. The unquestioning use of racial categories in scientific research can also lead to misplaced conclusions. For example, the use of race as a natural category can obfuscate the environmental, social, and structural factors that contribute to health disparities \cite{Sewell2016, Drevdahl2006} and lead to the racialization of certain diseases \cite{Foster2002, Duster1050, Braun2007}. The genomic era in particular has given rise to new concerns regarding the use of race in technologies and research. Race-based pharmaceuticals have been critiqued for their part in reifying racial categories as markers of biological difference \cite{kahn2005,Kelly2018}. Genetic ancestry testing has been criticized for its potential to promote biological essentialilsm and reinforce race privilege amongst those already experiencing it \cite{Roth2018, bolnick2007science}. Finally, bioinformatics tools and research practices themselves contribute to the essentialization of race \cite{Baren-Nawrocka2013, Fullwiley2007, Fullwiley2008, Helberg-Proctor2017}. 

There has also been significant debate about the appropriate use of racial variables in studies aimed at identifying causal effects \cite{gomez2013, james2008, Holland1986, holland2008, Kohler-Hausmann}. Objections to the use of race as a causal variable are often couched in terms of manipulability, embodied by the slogan "no causation without manipulation" \cite{Holland1986}. These arguments often point to the physical impossibility of manipulating race, thus precluding its use as a causal variable. Several additional, more sociologically-grounded objections to the use of race as a causal variable have emerged within social statistics \cite{holland2008}, anti-discrimination legal scholarship \cite{Kohler-Hausmann}, and health research \cite{gomez2013}. These objections point to the fundamental role race plays in structuring life experiences, making it nonsensical to talk about two individuals being identical, save for race. Stated another way, attempts to isolate the treatment effect of race are often based on a "sociologically incoherent conception of what race references" \cite{Kohler-Hausmann}. 

Several scholars have pointed to the significant social consequences that arise when race is misconstrued as a variable that can produce causal effects \cite{Zuberi2000, allen2008qui, gomez2013, james2008}. First, without proper contextualization of results, there is a tendency to misattribute the causal mechanisms of difference the racial categories themselves \cite{Zuberi2000, allen2008qui, gomez2013}, further reifying race as a natural category. Second, misidentifying race, rather than racial stratification, as the root cause of social and health disparities can lead to misplaced conclusions and ineffective public policy interventions \cite{gomez2013}.

\subsection{Critical race methodologies}

These critiques have given rise to a host of critical race methodologies informing the use of racial categories within these respective disciplines. To begin, the choice to use racial categories at all should be carefully examined, particularly in biological contexts. Despite widespread agreement that racial categories do not describe genetically distinct populations \cite{neal2008}, genomics researchers have grappled for over a decade with the utility and harm of using racial categories to frame research studies and communicate scientific results. The context of use is key to assessing the appropriate use of racial categories. For example, when adopted as a social or political category, race has utility in genomics, e.g. to document biological differences that result from processes of racial stratification. However, when disconnected from their social and political histories, racial categories have no place in biological research \cite{Phillips2008, Duster1050, Yudell564}. Genetics, biomedical researchers, and sociologists of science have emphasized the importance  of distinguishing ancestry from racial taxonomies within biomedical research \cite{Harawa2009, neal2008, fujimura2014clines} and have warned against the dangers of of using racial categories to describe genetic variation \cite{Yudell564, Duster1050, bliss2011}. 

As noted above in our discussion of Benthall and Haynes, race cannot and should not be done away with entirely. The key challenge here is to "denaturalize without dematerializing it" \cite{Mcharek2013} by recognizing race as a multidimensional, relational, and socially situated construct. This begins with centering the process of conceptualizing and operationalizing race, critically assessing the choice of categories and measurement schemes, and fully articulating and justifying these decisions in academic communications  \cite{Kaplan2003, gomez2013, mays2003, Lin2000, Morris2007}. 

In the context of causal studies, several methodologies have been proposed to mitigate the risks of reifying race as an entity that produces causal effects. For example,  some researchers have considered studies that manipulate properties associated with race, such as names \cite{Bertrand2004} \footnote{However, Kohler-Hausmann has been critical of such "audit" studies for not being able to isolate the treatment effect for race, although they do, in some circumstances, "provide evidence of a constitutive claim that grounds a thick ethical evaluation" \cite[pp. 33-34]{Kohler-Hausmann}}. Sen and Wasow's 'bundle of sticks' framework theorizes race as a composite variable that can be disaggregated into constitutive elements, some of which can be manipulated \cite{Sen2016}. 

Epidemiological and public health researchers have increasingly shifted their practices towards the study of social determinants of health, and in particular, the multiple ways racism affects health outcomes \cite{Fullilove1998, Kaufman2001, Williams2010, Williams1994, Williams2005, Lin2000}. This shift from studying effects of \textit{race} to the effects of \textit{racism} is a central component of critical race methodologies emerging within other disciplines. For instance, Ford and Hawara propose a methodology for the operationalization of race within health equity studies as a multidimensional, context-specific variable with a relational component to explicitly capture the effects of racial stratification on health outcomes \cite{Ford2010}.

Moreover, appropriately contextualized descriptive studies can be an important tool for identifying and understanding patterns of inequality. When race is utilized as a descriptive category, the choices about what categories to use and how to assign individuals to categories significantly impacts the results of analysis \cite{Saperstein2006, Bratter2011, Telles1998, Telles2015}. For example, Howell and Emerson compare the effectiveness of five different operationalizations of race in predicting different measures of social inequality. The study found each operationalization told a different story, pointing to the importance of critically evaluating racial categories and measurement schemes, as well as the importance of articulating these decisions in the communication of results \cite{Howell2016}.

In recent years, anti-racist efforts have emerged within public health that more thoroughly integrate critical race theory into methodologies and discourse \cite{Ford2010crt, ford2016}. While earlier studies of the social determinants of health tend to center individual and interpersonal racialized experiences, this new paradigm shifts focus to the institutional and structural conditions that shape racial disparities in health \cite{Sewell2016, Williams2001, Gee2005, García2015}. Anti-racist practice also emphasizes the importance of researchers recognizing their privilege and positionality and the way this might shape their practice. For example, the Public Health Critical Race Praxis offers a self-reflexive and race-conscious research methodology that centers discourse and practice around the perspectives of socially marginalized groups, building on community-based participatory approaches \cite{ford2016}.

\section{Implications for using race in algorithmic fairness research}


We observe several widespread tendencies when using race within algorithmic fairness research. First, frameworks for describing and mitigating unfairness adopt a simplistic conceptualization of race as a single dimensional variable that can take on a handful of values. This simplification erases the social, economic, and political complexity of the racial categories. Further, methodologies built upon this conceptualization of race tend to treat groups as interchangeable, obscuring the unique oppressions encountered by each group. We argue that this framing limits the effectiveness of fairness analysis and interventions and risks reifying racial categories in the process. 

Second, the process of conceptualizing and operationalizing race for the purposes of studying or mitigating different aspects of algorithmic fairness has -- with the exceptions noted above -- received little attention. Moreover, racial categories are frequently adopted with little attention given to the histories of the categories or suitability of the categories for the fairness assessment. 


We discuss the importance of focusing on categories and measurement processes, and fully articulating these decisions in communication of results. Lastly, we discuss the use of disaggregated analysis and the implications for incorporating a more nuanced understanding of measurement for these analyses.

\subsection{Race and group fairness}

Group fairness criteria represent a class of algorithmic fairness definitions predicated on clearly defined subgroups in the dataset. "Fairness" is obtained by equalizing a particular statistic, or set of statistics, of the classifier across groups. Many different group fairness criteria have been proposed. For example, demographic parity requires equal rates of positive prediction \cite{Feldman2015} across groups. Equality of odds \cite{Hardt2016}, also referred to as equalized mistreatment \cite{Zafar2017b}, requires equal false positive and false negative rates across groups. Equal opportunity \cite{Hardt2016} restricts this requirement to only a single value of the true outcome. Calibration (or test fairness) \cite{Chouldechova2017, Kleinberg2017} requires the actual outcome to be independent of protected attributes, conditioned on estimated outcome. 

While limitations of group fairness criteria have been previously explored \cite{Green2018TheMI, Corbett-Davies2018}, here we focus our examination on the conceptualization of race embedded within this framework. By abstracting racial categories into a mathematically comparable form, group-based fairness criteria deny the hierarchical nature and the social, economic, and political complexity of the social groups under consideration. Most notably, critical race theorists and Black feminist thinkers have criticized group fairness approaches for their underlying ideal, liberal approaches to ameliorating past harms.

Black feminist cultural geographer Katherine McKittrick, following Patricia Hill Collins, discusses how such approaches have resulted in a 'flattened geography' which obscures the unique oppressions encountered by Black women, instead treating oppressed social groups as interchangeable \cite{McKittrick2006, collins1998fighting}. Ladson-Billings and Tate, in their critical race critique of liberal education paradigms, explain that the tension between and even among different racial groups are not adequately examined or understood and assume "that all difference is both analogous and equivalent" \cite{ladson1995toward}. Political philosopher Charles W. Mills notes that the liberal underpinnings of both theoretical and methodological approaches towards equalization have historically tended to have significant detrimental consequences for racial minorities and only serves to advance a white "racial contract" \cite{mills1997racial}.

In short, group fairness approaches try to achieve sameness across groups without regard for the difference between the groups. Group fairness offers an incomplete version of what a race-conscious policy would be. This treats everyone the same from an algorithmic perspective without acknowledging that people are not treated the same. As political philosopher Elizabeth Anderson notes, "[s]tandard conceptions of distributive equality, embodied in ideas of equality of opportunity... fail to consider how such opportunities build in group hierarchy" \cite{anderson2009toward}.

\subsection{Conceptualizing and operationalizing race}

The question of how best to operationalize race for the purposes of studying or mitigating different aspects of algorithmic unfairness has received little attention. By discounting the considerations that go into operationalizing race, the scope and effectiveness of subsequent analysis and interventions fail to interrogate how the particular operationalization and measurement affect a given outcome. For instance, referring to Table \ref{tab:table1}, while observed race may be more important for studying discrimination, self-identification may be more suitable for studying identity formation and voting behavior.

We suggest centering the process of conceptualizing and operationalizing race when working with racial variables. 
In  particular, we urge algorithmic fairness researchers to critically evaluate existing racial schemas. Because of data limitations, we are most often bound to the categories provided by census categories or other taxonomies which stem from bureaucratic processes. We know from the histories outlined above that these categories are unstable, contingent, and rooted in racial inequality. 

Following recent work around measurement and fairness \cite{jacobs, andrus2019towards}, we suggest taking seriously the problems of measurement modeling when considering racial variables. Race, like fairness, is itself a contested concept, and it follows that adopting a multidimensional view is one strategy forward for approaching this. Recently, Roth has called more broadly for a "sociology of racial appraisals" in which we better theorize how observed race has changed over time, and how these changes influence norms of classification \cite{roth2018unsettled}. Given that most, if not all, of algorithmic fairness research is ostensibly concerned with anti-discrimination, this charge should be taken seriously.

As a helpful analogue, a growing body of survey research has highlighted the impact that racial category and measurement choices can have on outcomes being measured \cite{Saperstein2006, saperstein2012capturing, saperstein2016making}. Similarly, we emphasize that the various choices that go into the operationalization of race for the purposes of fairness-informed analysis or interventions significantly impact the result. Therefore, we need to be transparent about definitions, categories, measurements, and motivations for racial schema used in research and publication. If possible, multiple measures of race should be collected. In addition, measurement of race should be considered as an empirical problem in its own right. 


In domains where the categorization is based solely on observable characteristics, e.g. image datasets are annotated by third-parties, care needs to be taken to ensure race is not reduced to phenotype. This consideration can inform how the subgroups themselves are defined (e.g. shifting conceptually from racial categories to categories defined along phenotypic lines) and the way in which the dataset and analysis results are communicated with the larger research community (e.g. detailing procedures for determining categories and category membership, ensuring a distinction is made between phenotypically-determined categories and high-level racial categories). In the context of disaggregated evaluations in computer vision domains, the FAT* community has already seen a marked shift towards defining groups based on phenotypic properties rather than racial categories, thanks to the pioneering work of Buolamwini and colleagues. 


However, we also note that shifting to phenotypically-determined categories does not provide a full-stop solution. Given the dark history of physiognomy and phrenology \cite{y2017physiognomy}, in particular their role in promoting scientific racism and eugenics, researchers should be careful when imposing categories based on, for example, facial landmarks (e.g. IBM's diversity in faces dataset has labelled facial dimensions, proportions, etc.). Doing so risks devolving into using phenotypical features as inputs for a predictive models for individual characteristics and internal psychological states. Furthermore, while phenotype can be used as a means to understand processes of discrimination, practitioners {\it should not} equate race to phenotype, not to mention other biological markers such as genomes.

\subsection{Disaggregated analysis}

Frameworks for algorithmic audit studies \cite{Raji2019, buolamwini2018gender} and standardized model reporting guidelines \cite{Mitchell2019ModelCF} have highlighted the importance of reporting model statistics disaggregated by groups defined along cultural, demographic, or phenotypic lines. In contrast to group fairness definitions, which are often employed as ideals to be met, disaggregated analysis operates at a descriptive level. These analyses should begin from a pragmatist understanding of differences and interrogate the most salient aspects of race to be considered for particular technologies. For instance, in their audits of facial analysis technologies, Buolamwini and her colleagues concentrate on the phenotypical dimensions of race to understand the disparities of these vision-based systems\footnote{It should be noted that they do not take a similar approach towards gender.}. Results of disaggregated analysis depend heavily on the choice of categories and racial measurements used. 

Within survey research, sociologists of race have investigated the utility of different racial categories for the purposes of understanding markers of social inequality. Howell and Emerson find that a five-fold measure of self-identified race explains more variation in measures of inequality in income, housing, and health in the US \cite{Howell2016}. Saperstein and Penner find that racial categorization operates both as an input and an output to racial inequality. Individuals who experience an increase in social position (e.g. increased income) may be more likely to "lighten" themselves in survey responses, while those who become more disadvantaged (e.g. experiencing unemployment or incarceration) may "darken" themselves \cite{saperstein2012racial, penner2015disentangling}. The implication here is that it is necessary to understand the way that a particular racial dimension manifests in measurement and how it relates to the sociotechnical system under consideration. We encourage algorithmic fairness researchers to explore how different racial dimensions and their attendant measurements might reveal different patterns of unfairness in sociotechnical systems. 

\subsection{Limits of the algorithmic frame}

Before undertaking the project of systematizing social groups based on race, we should begin by interrogating why race is a relevant factor in the analysis of the system to begin with. Here, it is critical to expand the scope of analysis beyond the algorithmic frame \cite{Selbst2019} and interrogate how patterns of racial oppression might be embedded in the data and model and might interact with the resulting system. We urge researchers to question the following: is delineating data along racial lines critical to understanding patterns of fairness produced or reproduced by the system? 

In many cases, severe fairness concerns are evident prior to any quantitative race-based analysis of the system's outputs. For example, Rashida Richardson and colleagues recently published an extensive survey of the "dirty data" used to develop predictive policing systems in the US \cite{dirtydata}. The report reveals that in nine out of the thirteen jurisdictions reviewed, predictive policing systems ingested data that was generated while the jurisdiction was under investigation for corrupt, racially-biased, or otherwise illegal policing practices. This work represents an analysis centered not around \textit{race} but around \textit{racially disparate policing patterns}. This echoes much of the work from within health and biomedical sciences which emphasizes a racism effect rather than a race effect. 

\subsection{Centering perspectives of marginalized groups}

Lastly, interventions in the vein of algorithmic fairness need to consider the problems of the racially oppressed based on their view of injustices. Political philosopher Elizabeth Anderson proposes that we think about injustices from what she calls a \textit{non-ideal standpoint methodology} \cite{anderson2009toward}. The methodology is non-ideal insofar as that it views injustices from the point of view of the aggrieved groups and the particular harms these groups are facing. The methodology is "standpoint"-based insofar as we should begin from the perspectives of oppressed groups. Here, we can draw lessons from public health scholars who prioritize community-based participatory approaches. These  methodologies, which are grounded in critical race theory and feminist standpoint epistemology, take lived experiences of marginalized groups as a valuable and essential source of knowledge.
\section{Conclusion}

In this article, we have pulled back the frame from the algorithms involved in algorithmic fairness themselves and focused on the categories which constitute the "protected classes" used within these frameworks. We focus on race, given the critical nexus of race, technology, and inequality, and the prevalence of race in algorithmic fairness scholarship. We traced the histories of classification, eugenics, and state-sponsored category creation, and warned of the tendency to reify and naturalize racial categories which were the product of long histories of inequality. We then noted the multidimensionality of race and considered lessons from disciplines which have been dealing with this issue for decades. 
Lastly, we review the algorithmic fairness literature in light of these considerations. We highlight limitations of algorithmic fairness methodologies that adopt a simplistic and de-contextualized understanding of race and outline several suggestions for algorithmic fairness researchers moving forward.
Most in the FAT* community are what Ruha Benjamin calls the "overserved" \cite{benjamin2019race}. In this light, racial classifications need to be interrogated from the perspective of who they serve. Who is doing the categorizing and for what purpose? As in the beginning of this article, we echo J. Khadijah Abdurahman, "it is not just that classification systems are inaccurate or biased, it is who has the power to classify, to determine the repercussions / policies associated thereof and their relation to historical and accumulated injustice." Racial classifications can be an important tool in post-hoc disaggregation analysis. However, we need to know how to disaggregate, how to contextualize categories of disaggregation, and who that categorization serves.

\section*{Acknowledgements}

Thanks to Anna Lauren Hoffmann, Ben Hutchinson, Emma Kaywin, Issa Kohler-Haussman, Joan Fujimura, Lauren Mangels, and the three anonymous FAT* reviewers for helpful comments.

\bibliographystyle{plain}
\bibliography{references}

\begin{thebibliography}{100}

\bibitem{Abdurahman2019}
J.~Khadijah Abdurahman.
\newblock Fat* be wilin'.
\newblock 2019.

\bibitem{ali2016brief}
Syed~Mustafa Ali.
\newblock A brief introduction to decolonial computing.
\newblock {\em XRDS: Crossroads, The ACM Magazine for Students}, 22(4):16--21,
  2016.

\bibitem{allen2008qui}
Walter~R Allen, Susan~A Suh, Gloria Gonz{\'a}lez, and Joshua Yang.
\newblock Qui bono? explaining--or defending--winners and losers in the
  competition for educational achievement.
\newblock In Tukufu Zuberi and Eduardo Bonilla-Silva, editors, {\em White
  logic, white methods : racism and methodology}, chapter~13, pages 217--237.
  Rowman \& Littlefield Publishers, 2008.

\bibitem{anderson2009toward}
Elizabeth Anderson.
\newblock Toward a non-ideal, relational methodology for political philosophy:
  Comments on schwartzman's "challenging liberalism".
\newblock {\em Hypatia}, 24(4):130--145, 2009.

\bibitem{andrus2019towards}
McKane Andrus and Thomas~K Gilbert.
\newblock Towards a just theory of measurement: A principled social measurement
  assurance program for machine learning.
\newblock In {\em Proceedings of the 2019 AAAI/ACM Conference on AI, Ethics,
  and Society}, pages 445--451. ACM, 2019.

\bibitem{angwin2016j}
Julia Angwin and Jeff Larson.
\newblock Propublica responds to company's critique of machine bias story.
\newblock {\em ProPublica}, 29, 2016.

\bibitem{propublica}
Julia Angwin, Jeff Larson, Surya Mattu, and Lauren Kirchner.
\newblock Machine bias.
\newblock {\em ProPublica}, 2016.

\bibitem{Bailey2013}
Stanley~R. Bailey, Mara Loveman, and Jeronimo~O. Muniz.
\newblock Measures of "race" and the analysis of racial inequality in brazil.
\newblock {\em Social Science Research}, 42(1):106 -- 119, 2013.

\bibitem{barabas2017interventions}
Chelsea Barabas, Karthik Dinakar, Joichi Ito, Madars Virza, and Jonathan
  Zittrain.
\newblock Interventions over predictions: Reframing the ethical debate for
  actuarial risk assessment.
\newblock {\em arXiv preprint arXiv:1712.08238}, 2017.

\bibitem{Baren-Nawrocka2013}
Jan Baren-Nawrocka.
\newblock The bioinformatics of genetic origins: how identities become embedded
  in the tools and practices of bioinformatics.
\newblock {\em Life Sciences, Society and Policy}, 9:7, 09 2013.

\bibitem{benjamin2019race}
Ruha Benjamin.
\newblock {\em Race After Technology: Abolitionist Tools for the New Jim Code}.
\newblock John Wiley \& Sons, 2019.

\bibitem{Benthall2019}
Sebastian Benthall and Bruce~D. Haynes.
\newblock Racial categories in machine learning.
\newblock In {\em Proceedings of the Conference on Fairness, Accountability,
  and Transparency}, FAT* '19, 2019.

\bibitem{Bertrand2004}
Marianne Bertrand and Sendhil Mullainathan.
\newblock {Are Emily and Greg More Employable Than Lakisha and Jamal? A Field
  Experiment on Labor Market Discrimination}.
\newblock {\em American Economic Review}, 94(4):991--1013, September 2004.

\bibitem{bliss2011}
Catherine Bliss.
\newblock Racial taxonomy in genomics.
\newblock {\em Social Science \& Medicine}, 73(7):1019 -- 1027, 2011.

\bibitem{bolnick2007science}
Deborah~A Bolnick, Duana Fullwiley, Troy Duster, Richard~S Cooper, Joan~H
  Fujimura, Jonathan Kahn, Jay~S Kaufman, Jonathan Marks, Ann Morning, Alondra
  Nelson, et~al.
\newblock The science and business of genetic ancestry testing.
\newblock {\em Science}, 318(5849):399--400, 2007.

\bibitem{bonilla2008toward}
Eduardo Bonilla-Silva and Tukufu Zuberi.
\newblock Toward a definition of white logic and white methods.
\newblock In Tukufu Zuberi and Eduardo Bonilla-Silva, editors, {\em White
  logic, white methods: racism and methodology}, chapter~1, pages 3--27. Rowman
  \& Littlefield Publishers, 2008.

\bibitem{sorting}
Geoffrey~C. Bowker and Susan~Leigh Star.
\newblock {\em Sorting Things Out}.
\newblock MIT Press, 2000.

\bibitem{Bratter2011}
Jenifer Bratter and Bridget K~Gorman.
\newblock Does multiracial matter? a study of racial disparities in self-rated
  health.
\newblock {\em Demography}, 48:127--52, 02 2011.

\bibitem{Braun2007}
Lundy Braun, Anne Fausto-Sterling, Duana Fullwiley, Evelynn Hammonds, Alondra
  Nelson, William Quivers, Susan Reverby, and Alexandra E~Shields.
\newblock Racial categories in medical practice: How useful are they?
\newblock {\em PLoS medicine}, 4:e271, 10 2007.

\bibitem{brock2009life}
Andre Brock.
\newblock Life on the wire: Deconstructing race on the internet.
\newblock {\em Information, Communication \& Society}, 12(3):344--363, 2009.

\bibitem{browne2015dark}
Simone Browne.
\newblock {\em Dark matters: On the surveillance of blackness}.
\newblock Duke University Press, 2015.

\bibitem{buolamwini2018gender}
Joy Buolamwini and Timnit Gebru.
\newblock Gender shades: Intersectional accuracy disparities in commercial
  gender classification.
\newblock In {\em FAT*}, pages 77--91, 2018.

\bibitem{Chouldechova2017}
Alexandra Chouldechova.
\newblock Fair prediction with disparate impact: A study of bias in recidivism
  prediction instruments.
\newblock {\em Big Data}, 5 2:153--163, 2017.

\bibitem{chun2009}
Wendy Hui~Kyong Chun.
\newblock {Introduction: Race and/as Technology; or, How to Do Things to Race}.
\newblock {\em Camera Obscura: Feminism, Culture, and Media Studies}, 24(1
  (70)):7--35, 05 2009.

\bibitem{collins1998fighting}
Patricia~Hill Collins.
\newblock {\em Fighting words: Black women and the search for justice},
  volume~7.
\newblock U of Minnesota Press, 1998.

\bibitem{Corbett-Davies2018}
Sam Corbett-Davies and Sharad Goel.
\newblock The measure and mismeasure of fairness: A critical review of fair
  machine learning.
\newblock In {\em arXiv:1808.00023}, 2018.

\bibitem{desmond2009racial}
Matthew Desmond and Mustafa Emirbayer.
\newblock What is racial domination?
\newblock {\em Du Bois Review: Social Science Research on Race}, 6(2):335--355,
  2009.

\bibitem{dieterich2016compas}
William Dieterich, Christina Mendoza, and Tim Brennan.
\newblock {COMPAS risk scales: Demonstrating accuracy equity and predictive
  parity}.
\newblock {\em Northpoint Inc}, 2016.

\bibitem{Drevdahl2006}
Denise~J. Drevdahl, Debby~A. Philips, and Janette~Y. Taylor.
\newblock Uncontested categories: the use of race and ethnicity variables in
  nursing research.
\newblock {\em Nursing Inquiry}, 13(1):52--63, 2006.

\bibitem{Duster1050}
Troy Duster.
\newblock Race and reification in science.
\newblock {\em Science}, 307(5712):1050--1051, 2005.

\bibitem{elder2012towards}
Dave Elder-Vass.
\newblock Towards a realist social constructionism.
\newblock {\em Sociologia, problemas e pr{\'a}ticas}, (70):9--24, 2012.

\bibitem{elliott2019assessing}
Diana Elliott, Rob Santos, Steven Martin, and Charmaine Runes.
\newblock Assessing miscounts in the 2020 census.
\newblock 2019.

\bibitem{Feldman2015}
Michael Feldman, Sorelle~A. Friedler, John Moeller, Carlos Scheidegger, and
  Suresh Venkatasubramanian.
\newblock Certifying and removing disparate impact.
\newblock In {\em Proceedings of the 21th ACM SIGKDD International Conference
  on Knowledge Discovery and Data Mining}, KDD '15, pages 259--268, New York,
  NY, USA, 2015. ACM.

\bibitem{Ford2010}
Chandra Ford and Nina Harawa.
\newblock A new conceptualization of ethnicity for social epidemiologic and
  health equity research.
\newblock {\em Social science \& Medicine}, 71:251--8, 07 2010.

\bibitem{ford2016}
Chandra~L. Ford.
\newblock Public health critical race praxis: An introduction, an intervention,
  and three points for consideration.
\newblock {\em Wisconsin Law Review}, 3:477--491, 01 2016.

\bibitem{Ford2010crt}
Chandra~L. Ford and Collins Airhihenbuwa.
\newblock Critical race theory, race equity, and public health: Toward
  antiracism praxis.
\newblock {\em American Journal of Public Health}, 100 Suppl 1:S30--5, 02 2010.

\bibitem{Foster2002}
Morris~W. Foster and Richard~R Sharp.
\newblock Race, ethnicity, and genomics: social classifications as proxies of
  biological heterogeneity.
\newblock {\em Genome research}, 12 6:844--50, 2002.

\bibitem{foucault2005order}
Michel Foucault.
\newblock {\em The order of things}.
\newblock Routledge, 2005.

\bibitem{fourcade2016seeing}
Marion Fourcade and Kieran Healy.
\newblock Seeing like a market.
\newblock {\em Socio-Economic Review}, 15(1):9--29, 2016.

\bibitem{fujimura2014clines}
Joan~H Fujimura, Deborah~A Bolnick, Ramya Rajagopalan, Jay~S Kaufman, Richard~C
  Lewontin, Troy Duster, Pilar Ossorio, and Jonathan Marks.
\newblock Clines without classes: How to make sense of human variation.
\newblock {\em Sociological Theory}, 32(3):208--227, 2014.

\bibitem{Fullilove1998}
Mindy Fullilove.
\newblock Comment: Abandoning "race" as a variable in public health research -
  an idea whose time has come.
\newblock {\em American Journal of Public Health}, 88:1297--8, 10 1998.

\bibitem{Fullwiley2007}
Duana Fullwiley.
\newblock The molecularization of race: Institutionalizing human difference in
  pharmacogenetics practice.
\newblock {\em Science as Culture}, 16(1):1--30, 2007.

\bibitem{Fullwiley2008}
Duana Fullwiley.
\newblock The biologistical construction of race.
\newblock {\em Social studies of science}, 38:695--735, 11 2008.

\bibitem{García2015}
Jennifer Garc{\`i}a and Mienah Sharif.
\newblock Black lives matter: A commentary on racism and public health.
\newblock {\em American Journal of Public Health}, 105:e1--e4, 06 2015.

\bibitem{Gee2005}
Gilbert Gee and Devon Payne-Sturges.
\newblock Environmental health disparities: A framework integrating
  psychosocial and environmental concepts.
\newblock {\em Environmental health perspectives}, 112:1645--53, 01 2005.

\bibitem{gomez2013}
Laura~E. Gomez and Nancy Lopez, editors.
\newblock {\em Mapping Race: Critical Approaches to Health Disparities
  Research}.
\newblock New Brunswick, NJ, Rutgers University Press, 2013.

\bibitem{Green2018TheMI}
Ben Green and Lily Hu.
\newblock The myth in the methodology: Towards a recontextualization of
  fairness in machine learning.
\newblock In {\em ICML 2018}, 2018.

\bibitem{hacking1999social}
Ian Hacking.
\newblock {\em The social construction of what?}
\newblock Harvard university press, 1999.

\bibitem{Harawa2009}
Nina~T Harawa and C.~Lawrence Ford.
\newblock The foundation of modern racial categories and implications for
  research on black/white disparities in health.
\newblock {\em Ethnicity \& Disease}, 19 2:209--17, 2009.

\bibitem{Hardt2016}
Moritz Hardt, Eric Price, and Nathan Srebro.
\newblock Equality of opportunity in supervised learning.
\newblock In {\em Proceedings of the 30th International Conference on Neural
  Information Processing Systems}, NIPS'16, pages 3323--3331, USA, 2016. Curran
  Associates Inc.

\bibitem{Helberg-Proctor2017}
Alana Helberg-Proctor, Anja Krumeich, Agnes Meershoek, and Klasien Horstman.
\newblock The multiplicity and situationality of enacting 'ethnicity' in dutch
  health research articles.
\newblock {\em BioSocieties}, 12 2017.

\bibitem{herzfeld1993social}
Michael Herzfeld.
\newblock {\em The social production of indifference}.
\newblock University of Chicago Press, 1993.

\bibitem{hoffmann2019fairness}
Anna~Lauren Hoffmann.
\newblock Where fairness fails: data, algorithms, and the limits of
  antidiscrimination discourse.
\newblock {\em Information, Communication \& Society}, 22(7):900--915, 2019.

\bibitem{Holland1986}
Paul~W. Holland.
\newblock Statistics and causal inference.
\newblock {\em Journal of the American Statistical Association},
  81(396):945--960, 1986.

\bibitem{holland2008}
Paul~W. Holland.
\newblock Causation and race.
\newblock In Tukufu Zuberi and Eduardo Bonilla-Silva, editors, {\em White
  logic, white methods : racism and methodology}, chapter~5, pages 93--109.
  Rowman \& Littlefield Publishers, 2008.

\bibitem{Howell2016}
Junia Howell and Michael Emerson.
\newblock So what "should" we use? evaluating the impact of five racial
  measures on markers of social inequality.
\newblock {\em Sociology of Race and Ethnicity}, 3:14--30, 5 2017.

\bibitem{jacobs}
Abigail~Z. Jacobs and Hanna Wallach.
\newblock Measurement and fairness.
\newblock 2019.

\bibitem{james2008}
Angela James.
\newblock Making sense of race and racial classification.
\newblock In Tukufu Zuberi and Eduardo Bonilla-Silva, editors, {\em White
  logic, white methods : racism and methodology}, chapter~2, pages 31--45.
  Rowman \& Littlefield Publishers, 2008.

\bibitem{kahn2005}
Jonathan Kahn.
\newblock Misreading race and genomics after bidil.
\newblock {\em Nature genetics}, 37:655--6, 08 2005.

\bibitem{Kaplan2003}
Judith~B. Kaplan and Trude Bennett.
\newblock {Use of Race and Ethnicity in Biomedical Publication}.
\newblock {\em JAMA}, 289(20):2709--2716, 2003.

\bibitem{Kaufman2001}
Jay Kaufman and Richard Cooper.
\newblock Commentary: Considerations for use of racial/ethnic classification in
  etiologic research.
\newblock {\em American journal of epidemiology}, 154:291--8, 09 2001.

\bibitem{Kaufman1999}
Jay~S Kaufman.
\newblock How inconsistencies in racial classification demystify the race
  construct in public health statistics.
\newblock {\em Epidemiology}, 10 2:101--3, 1999.

\bibitem{Kelly2018}
Shannon Kelly and Yashwant Pathak.
\newblock {\em Race and Ethnicity: Understanding Difference in the Genome Era},
  pages 71--87.
\newblock 07 2018.

\bibitem{kendi2017stamped}
Ibram~X Kendi.
\newblock {\em Stamped from the beginning: The definitive history of racist
  ideas in America}.
\newblock Random House, 2017.

\bibitem{khalfani2008race}
Akil~Kokayi Khalfani, Tukufu Zuberi, Sulaiman Bah, and Pali~J Lehohla.
\newblock Race and population statistics in south africa.
\newblock In Tukufu Zuberi and Eduardo Bonilla-Silva, editors, {\em White
  logic, white methods: racism and methodology}, chapter~4, pages 63--92.
  Rowman \& Littlefield Publishers, 2008.

\bibitem{Kleinberg2017}
Jon~M. Kleinberg, Sendhil Mullainathan, and Manish Raghavan.
\newblock Inherent trade-offs in the fair determination of risk scores.
\newblock In {\em ITCS}, 2017.

\bibitem{Kohler-Hausmann}
Issa Kohler-Hausmann.
\newblock {Eddie Murphy and the Dangers of Counterfactual Causal Thinking About
  Detecting Racial Discrimination}.
\newblock {\em Northwestern University Law Review}, 113(5), 2019.

\bibitem{ladson1995toward}
Gloria Ladson-Billings and William~F Tate~IV.
\newblock Toward a critical race theory of education.
\newblock {\em Teachers College Record}, 97:47--68, 1995.

\bibitem{larson2016we}
Jeff Larson, Surya Mattu, Lauren Kirchner, and Julia Angwin.
\newblock How we analyzed the compas recidivism algorithm.
\newblock {\em ProPublica}, 9, 2016.

\bibitem{Lee2009}
Catherine Lee.
\newblock "race" and "ethnicity" in biomedical research: How do scientists
  construct and explain differences in health?
\newblock {\em Social Science \& Medicine}, 68(6):1183 -- 1190, 2009.

\bibitem{Lin2000}
S~Lin and J~Kelsey.
\newblock Use of race and ethnicity in epidemiologic research: Concepts,
  methodological issues, and suggestions for research.
\newblock {\em Epidemiologic reviews}, 22:187--202, 02 2000.

\bibitem{liptak2019}
Adam Liptak.
\newblock Supreme court leaves census question on citizenship in doubt.
\newblock {\em The New York Times}, 2019.

\bibitem{lopez2006white}
Ian~Haney L{\'o}pez.
\newblock {\em White by law: The legal construction of race}.
\newblock NYU Press, 2006.

\bibitem{Loveman2012}
Mara Loveman, Jeronimo Muniz, and Stanley R.~Bailey.
\newblock Brazil in black and white? race categories, the census, and the study
  of inequality.
\newblock {\em Ethnic and Racial Studies}, 35, 08 2012.

\bibitem{maghbouleh2017limits}
Neda Maghbouleh.
\newblock {\em The Limits of Whiteness: Iranian Americans and the Everyday
  Politics of Race}.
\newblock Stanford University Press, 2017.

\bibitem{massey1993american}
Douglas~S Massey and Nancy~A Denton.
\newblock {\em American apartheid: Segregation and the making of the
  underclass}.
\newblock Harvard University Press, 1993.

\bibitem{mays2003}
Vickie Mays, Ninez Ponce, Donna Washington, and Susan Cochran.
\newblock Classification of race and ethnicity: Implications for public health.
\newblock {\em Annual Review of Public Health}, 24:83--110, 02 2003.

\bibitem{Mcharek2013}
Amade M'charek.
\newblock Beyond fact or fiction: On the materiality of race in practice.
\newblock {\em Cultural Anthropology}, 28:420--442, 07 2013.

\bibitem{McKittrick2006}
Katherine McKittrick.
\newblock {\em Demonic Grounds: Black Women and the Cartographies of Struggle}.
\newblock University of Minnesota Press, ned - new edition edition, 2006.

\bibitem{mills1997racial}
Charles~W Mills.
\newblock {\em The racial contract}.
\newblock Cornell University Press, 1997.

\bibitem{Mitchell2019ModelCF}
Margaret Mitchell, Simone Wu, Andrew Zaldivar, Parker Barnes, Lucy Vasserman,
  Ben Hutchinson, Elena Spitzer, Inioluwa~Deborah Raji, and Timnit Gebru.
\newblock Model cards for model reporting.
\newblock In {\em FAT}, 2019.

\bibitem{monk2015cost}
Ellis~P Monk~Jr.
\newblock The cost of color: Skin color, discrimination, and health among
  african-americans.
\newblock {\em American Journal of Sociology}, 121(2):396--444, 2015.

\bibitem{Morris2007}
Edward Morris.
\newblock Researching race: Identifying a social construction through
  qualitative methods and an interactionist perspective.
\newblock {\em Symbolic Interaction}, 30:409--425, 08 2007.

\bibitem{neal2008}
Karama~C. Neal.
\newblock {Use and Misuse of 'Race' in Biomedical Research}.
\newblock {\em Online Journal of Health Ethics}, 5(1), 2008.

\bibitem{noble2018algorithms}
Safiya~Umoja Noble.
\newblock {\em Algorithms of oppression: How search engines reinforce racism}.
\newblock nyu Press, 2018.

\bibitem{omi2014racial}
Michael Omi and Howard Winant.
\newblock {\em Racial formation in the United States}.
\newblock Routledge, 2014.

\bibitem{omi2001changing}
Michael~A Omi.
\newblock The changing meaning of race.
\newblock {\em America becoming: Racial trends and their consequences},
  1:243--263, 2001.

\bibitem{pascoe1996miscegenation}
Peggy Pascoe.
\newblock Miscegenation law, court cases, and ideologies of "race" in
  twentieth-century america.
\newblock {\em The Journal of American History}, 83(1):44--69, 1996.

\bibitem{penner2015disentangling}
Andrew~M Penner and Aliya Saperstein.
\newblock Disentangling the effects of racial self-identification and
  classification by others: the case of arrest.
\newblock {\em Demography}, 52(3):1017--1024, 2015.

\bibitem{Phillips2008}
Elizabeth Phillips, Adebola Odunlami, and Vence Bonham.
\newblock Mixed race: Understanding difference in the genome era.
\newblock {\em Social forces; a scientific medium of social study and
  interpretation}, 86:795--820, 01 2008.

\bibitem{rajagopalan2012making}
Ramya Rajagopalan and Joan~H Fujimura.
\newblock Making history via dna, making dna from history.
\newblock {\em Genetics and the unsettled past: The collision of DNA, race, and
  history}, page 143, 2012.

\bibitem{Raji2019}
Inioluwa~Deborah Raji and Joy Buolamwini.
\newblock Actionable auditing: Investigating the impact of publicly naming
  biased performance results of commercial ai products.
\newblock In {\em AIES}, 2019.

\bibitem{ray2019theory}
Victor Ray.
\newblock A theory of racialized organizations.
\newblock {\em American Sociological Review}, 84(1):26--53, 2019.

\bibitem{dirtydata}
Rashida Richardson, Jason Schultz, and Kate Crawford.
\newblock Dirty data, bad predictions: How civil rights violations impact
  police data, predictive policing systems, and justice.
\newblock {\em New York University Law Review}, 2019.

\bibitem{roth2016multiple}
Wendy~D Roth.
\newblock The multiple dimensions of race.
\newblock {\em Ethnic and Racial Studies}, 39(8):1310--1338, 2016.

\bibitem{Roth2017}
Wendy~D. Roth.
\newblock Methodological pitfalls of measuring race: international comparisons
  and repurposing of statistical categories.
\newblock {\em Ethnic and Racial Studies}, 40(13):2347--2353, 2017.

\bibitem{roth2018unsettled}
Wendy~D Roth.
\newblock Unsettled identities amid settled classifications? toward a sociology
  of racial appraisals.
\newblock {\em Ethnic and Racial Studies}, 41(6):1093--1112, 2018.

\bibitem{Roth2018}
Wendy~D. Roth and Biorn Ivemark.
\newblock Genetic options: The impact of genetic ancestry testing on consumers'
  racial and ethnic identities.
\newblock {\em American Journal of Sociology}, 124(1):150--184, 2018.

\bibitem{rugh2010racial}
Jacob~S Rugh and Douglas~S Massey.
\newblock Racial segregation and the american foreclosure crisis.
\newblock {\em American sociological review}, 75(5):629--651, 2010.

\bibitem{Salter2013}
Phia Salter and Glenn Adams.
\newblock Toward a critical race psychology.
\newblock {\em Social and Personality Psychology Compass}, 7, 11 2013.

\bibitem{Sandefur2004}
Gary~D. Sandefur, Mary~E. Campbell, and Jennifer Eggerling-Boeck.
\newblock Racial and ethnic disparities in health and mortality among the u.s.
  elderly population.
\newblock In Anderson NB, Bulatao RA, and Cohen B, editors, {\em Critical
  Perspectives on Racial and Ethnic Differences in Health in Late Life}, pages
  53--94. Washington, DC: The National Academies Press, 2004.

\bibitem{Saperstein2006}
Aliya Saperstein.
\newblock Double-checking the race box: Examining inconsistency between survey
  measures of observed and self-reported race.
\newblock {\em Social Forces}, 85(1):57--74, 2006.

\bibitem{saperstein2012capturing}
Aliya Saperstein.
\newblock Capturing complexity in the united states: which aspects of race
  matter and when?
\newblock {\em Ethnic and Racial Studies}, 35(8):1484--1502, 2012.

\bibitem{saperstein2016making}
Aliya Saperstein, Jessica~M Kizer, and Andrew~M Penner.
\newblock Making the most of multiple measures: Disentangling the effects of
  different dimensions of race in survey research.
\newblock {\em American Behavioral Scientist}, 60(4):519--537, 2016.

\bibitem{saperstein2012racial}
Aliya Saperstein and Andrew~M Penner.
\newblock Racial fluidity and inequality in the united states.
\newblock {\em American Journal of Sociology}, 118(3):676--727, 2012.

\bibitem{scott1998seeing}
James~C Scott.
\newblock {\em Seeing like a state: How certain schemes to improve the human
  condition have failed}.
\newblock Yale University Press, 1998.

\bibitem{scott2017against}
James~C Scott.
\newblock {\em Against the grain: a deep history of the earliest states}.
\newblock Yale University Press, 2017.

\bibitem{Selbst2019}
Andrew~D. Selbst, danah boyd, Sorelle~A. Friedler, Suresh Venkatasubramanian,
  and Janet Vertesi.
\newblock Fairness and abstraction in sociotechnical systems.
\newblock In {\em Proceedings of the Conference on Fairness, Accountability,
  and Transparency}, FAT* '19, pages 59--68, New York, NY, USA, 2019. ACM.

\bibitem{Sen2016}
Maya Sen and Omar Wasow.
\newblock Race as a bundle of sticks: Designs that estimate effects of
  seemingly immutable characteristics.
\newblock {\em Annual Review of Political Science}, 19:499--522, 05 2016.

\bibitem{Sewell2016}
Abigail~A. Sewell.
\newblock The racism-race reification process: A mesolevel political economic
  framework for understanding racial health disparities.
\newblock {\em Sociology of Race and Ethnicity}, 2(4):402--432, 2016.

\bibitem{smart2008}
Andrew Smart, Richard Tutton, Paul Martin, George~T.H. Ellison, and Richard
  Ashcroft.
\newblock The standardization of race and ethnicity in biomedical science
  editorials and uk biobanks.
\newblock {\em Social Studies of Science}, 38(3):407--423, 2008.

\bibitem{snipp2003racial}
C~Matthew Snipp.
\newblock Racial measurement in the american census: Past practices and
  implications for the future.
\newblock {\em Annual Review of Sociology}, 29(1):563--588, 2003.

\bibitem{tallbear2013native}
Kim TallBear.
\newblock {\em Native American DNA: Tribal belonging and the false promise of
  genetic science}.
\newblock U of Minnesota Press, 2013.

\bibitem{Telles2015}
Edward Telles, Ren{\'e}~D. Flores, and Fernando Urrea-Giraldo.
\newblock Pigmentocracies: Educational inequality, skin color and census
  ethnoracial identification in eight latin american countries.
\newblock {\em Research in Social Stratification and Mobility}, 40:39 -- 58,
  2015.

\bibitem{Telles1998}
Edward~E. Telles and Nelson Lim.
\newblock Does it matter who answers the race question? racial classification
  and income inequality in brazil.
\newblock {\em Demography}, 35(4):465--474, 1998.

\bibitem{tilly1990coercion}
Charles Tilly.
\newblock {\em Coercion, Capital, and European States, AD 990-1992}.
\newblock B. Blackwell, 1990.

\bibitem{Williams2005}
David Williams and Pamela Braboy~Jackson.
\newblock Social sources of racial disparities in health.
\newblock {\em Health affairs (Project Hope)}, 24:325--34, 03 2005.

\bibitem{Williams2001}
David Williams and Chiquita Collins.
\newblock Racial residential segregation: A fundamental cause of racial
  disparities in health.
\newblock {\em Public Health Reports}, 116:404--416, 09 2001.

\bibitem{Williams1994}
David~R. Williams.
\newblock The concept of race in health services research: 1966 to 1990.
\newblock {\em Health Services Research}, 29(3):261--274, 1994.

\bibitem{Williams2010}
David~R. Williams and Michelle Sternthal.
\newblock Understanding racial-ethnic disparities in health: sociological
  contributions.
\newblock {\em Journal of Health and Social Behavior}, 51
  Suppl(Suppl):S15--S27, 2010.

\bibitem{y2017physiognomy}
Blaise~Ag{\"u}era y~Arcas, Margaret Mitchell, and Alexander Todorov.
\newblock Physiognomy's new clothes.
\newblock {\em Medium}, 2017.

\bibitem{Yudell564}
Michael Yudell, Dorothy Roberts, Rob DeSalle, and Sarah Tishkoff.
\newblock Taking race out of human genetics.
\newblock {\em Science}, 351(6273):564--565, 2016.

\bibitem{Zafar2017b}
Muhammad~Bilal Zafar, Isabel Valera, Manuel Gomez~Rodriguez, and Krishna~P.
  Gummadi.
\newblock Fairness beyond disparate treatment \& disparate impact: Learning
  classification without disparate mistreatment.
\newblock In {\em Proceedings of the 26th International Conference on World
  Wide Web}, WWW '17, pages 1171--1180, Republic and Canton of Geneva,
  Switzerland, 2017. International World Wide Web Conferences Steering
  Committee.

\bibitem{Zuberi2000}
Tukufu Zuberi.
\newblock Deracializing social statistics: Problems in the quantification of
  race.
\newblock {\em The Annals of the American Academy of Political and Social
  Science}, 568:172--185, 2000.

\bibitem{zuberi2001thicker}
Tukufu Zuberi.
\newblock {\em Thicker Than Blood: How Racial Statistics Lie}.
\newblock U of Minnesota Press, 2001.

\end{thebibliography}

\end{document}